\documentstyle[aps,prb,preprint]{revtex}
\tightenlines
\begin{document}
\draft

\title{Spin-polarized tunneling of\\
La$_{0.67}$Sr$_{0.33}$MnO$_3$/YBa$_2$Cu$_3$O$_{7-\delta}$
junctions}

\author{A. Sawa, S. Kashiwaya, H. Obara, H. Yamasaki, and M. Koyanagi}
\address{Electrotechnical Laboratory, 1-1-4 Umezono,
Tsukuba, Ibaraki 305-8568, Japan.}

\author{N. Yoshida and Y. Tanaka}
\address{Department of Applied Physics, Nagoya University, 464-8603,
Nagoya, Japan.}

\date{\today}
\maketitle

\begin{abstract}

The transport properties
between ferromagnets and high-$T_{c}$ superconductors were investigated in La$_{0.67}$Sr$_{0.33}$MnO$_3$/YBa$_2$Cu$_3$O$_{7-\delta}$ (LSMO/YBCO) junctions in the geometry of cross-strip lines.
The conductance spectra show zero-bias conductance peaks (ZBCP),
reflecting the charge transport in the $ab$-plane.
When an external magnetic field is applied to the junctions,
the conductance spectra show two notable features, $i$.$e$., 
an increase of background conductance
and an asymmetric ZBCP splitting whose amplitude responds nonlinearly to the applied field.
It is shown that the magnetic field response are consistent with a theoretical prediction of tunneling spectroscopy when the presence of a ferromagnetic barrier between a spin-polarized ferromagnet and a $d$-wave superconductor is assumed.

\end{abstract}



\section{Introduction}
Hybrid structures between
ferromagnets and superconductors
have been the focus of much attention
in terms of spin-dependent spectroscopy
and spin-injection devices.
The fundamental properties of
ferromagnet/insulator/superconductor (F/I/S) junctions
fabricated from conventional
metal superconductors
have been studied
for about 30 years \cite{meservey_1}.
The recent rediscovery of perovskite manganites which exhibit colossal magnetoresistance (CMR) has aroused a new possibility in this field,
because the layered structure fabrication of ferromagnets and high-$T_c$
superconductors is possible
using these oxide compounds
\cite{chahara_1,helmolt_1,vasko_1,dong_1,vasko_2,yeh}.
There are two main intriguing aspects in this field.
One is the influence of the carrier injection on
bulk superconducting properties,
such as the suppression of the critical current density
and the critical
temperature due to the nonequilibrium states \cite{tinkham,clarke}.
The injection of spin-polarized quasiparticles
is expected to enhance the nonequiliburium
because the spin-relaxation time is estimated to be
much longer than
the quasiparticle-recombination time
in light metals \cite{johnson}.
It has been experimentally verified that suppression of the critical current  in high-$T_c$ superconductors is induced due to a spin-polarized quasiparticle injection from ferromagnets of CMR compounds \cite{vasko_1,dong_1,yeh} or pure metals \cite{iguchi}.
The other aspect is the boundary properties,
such as the connection of wave functions,
the Andreev reflection \cite{andreev,blonder},
and bound states formation
at the surface for $p$-wave \cite{buch,hara}
and for $d$-wave superconductors
\cite{hu,tanaka_1,tanaka_2,kashiwaya_2,kashiwaya_3}.
Moreover, several theories have
elucidated the transport properties
under the influence of an exchange field
for $s$-wave superconductors \cite{deJong}
and for $d$-wave superconductors
\cite{zhu_1,zutic_1,kashiwaya_1,yoshida}.
However, detailed comparisons between
theory and experiment
have not yet been accomplished.
\par
In this paper, we study the transport properties and
magnetic field response of La$_{0.67}$Sr$_{0.33}$MnO$_3$/YBa$_2$Cu$_3$O$_{7-\delta}$ (LSMO/YBCO) cross-strip type junctions.
We discuss the properties of the interface between ferromagnets and high-$T_c$ superconductors, which is important even for the injection device
because the spin injection utilizes transport through surfaces.
In the case of $d$-wave superconductors,
the formation of surface bound states is known to modify
the transport property \cite{tanaka_1,kashiwaya_2}.
When the orientation of the junction is along the $ab$-plane
and the misorientation of the $a$-axis to the boundary
is finite, zero-energy bound states are formed at the
boundary \cite{hu}.
The presence of the zero-energy states at the boundary of
high-$T_c$ superconductors has been
detected as zero-bias conductance peaks (ZBCPs) in
a wide variety of tunneling junctions
\cite{tanaka_1,kashiwaya_2,lesueur_1,alff_1,covington_1,wei,sanders_1}.
The modification of this property in LSMO/YBCO junctions is an interesting problem,
where quasiparticles in the normal side are strongly
spin polarized.
This is because the transport properties of the junctions are expected to
be sensitive to an applied magnetic field
and to the interface properties.

\section{Experimental}
\label{exp}
%
Because of their simple geometry,
cross-strip junctions are used for measurements.
Figure \ref{fig1} shows schematic illustrations
of the top view and  the cross-sectional view of
the junctions.
$C$-axis oriented
epitaxial YBCO thin films  of 100 to 150
nm thickness were deposited on
SrTiO$_3$ (100) substrates by pulsed laser deposition (PLD).
The films were patterned into bridges 30--60 $\mu$m wide and 30--100
$\mu$m long by conventional photolithography and wet chemical etching employing phosphoric acid.
Next, 100-nm-thick LSMO films were deposited on the patterned YBCO films,
also by PLD.
After the deposition, the films were subsequently annealed at 400 $^{\circ}$C in an oxygen atmosphere for 1--2 hours.
The LSMO/YBCO layered films were patterned into cross-strip structures by Ar ion milling.
The substrate temperature was 750 $^{\circ}$C for the deposition of YBCO,
and 700 $^{\circ}$C for LSMO.
The laser energy density was 1.5 J/cm$^2$ for YBCO, and 2 J/cm$^2$ for LSMO.
The laser repetition frequency was 2 Hz and the oxygen pressure was 700 mTorr for both materials.
Au film contact pads, annealed at 400 $^{\circ}$C in an oxygen atmosphere
to reduce the contact resistance, were used as electrodes.

LSMO films on the $c$-axis-oriented YBCO films were confirmed to be $c$-axis oriented by X-ray diffraction measurement.
Temperature dependence of resistance showed that the
superconducting transition
temperature $T_c$ of YBCO was about 90 K,
and the magnetization measurement showed that the ferromagnetic transition
temperature of LSMO was about 350 K.
\par
The $I$-$V$ characteristics of the junctions were measured using a dc
four-probe method and the conductance spectra ($i.e.$, $dI/dV$-$V$ curves) were numerically calculated from the $I$-$V$ data.
The magnetic fields of 0--12 T generated by a superconducting magnet were
applied along the direction parallel to the film surface and perpendicular
to the trajectories of tunneling electrons.
\par
%
\section{Results and Discussion}
\label{result}
%
In the following, experimental results of conductance spectra and their magnetic field response in the LSMO/YBCO junctions are presented.
Since most of our samples exhibited similar features,
we concentrate on the data obtained from the sample with the junction area of $30\times30$ $\mu$m$^2$.
We will demonstrate two features
peculiar to the LSMO/YBCO junctions:
the tunneling electron is actually spin-polarized,
and the barrier naturally formed between YBCO and LSMO
behaves as a ferromagnetic insulator,
leading to the spin-filtering effect.
The conductance spectra
are analyzed based on a theoretical formula
for ferromagnet/ferromagnetic insulator/$d$-wave superconductor
(F/FI/d) junctions and the notation used in the analysis mostly follows that used in Ref. \cite{kashiwaya_1}.
A cylindrical Fermi surface is assumed with the Fermi energy $E_{FS}$ of 0.3 eV in YBCO,
and the effective masses are set to be equal in YBCO and LSMO.
For the model of ferromagnet, the Stoner model is adopted.
The polarization $P$ and the Fermi-wave vector of quasiparticles for up[down]-spins $k_{N,\uparrow[\downarrow]}$ in LSMO are not independent parameters in the framework of the Stoner model \cite{note_1}.
The normalized barrier heights for up[down]-spin $V_{\uparrow[\downarrow]}$ in a ferromagnetic insulator are used as the fitting parameters in the following analysis.
We note that the formula used in the present study has some drawbacks in the analysis of an actual junction.
One is that the delta function form $V_{\uparrow[\downarrow]}\delta(x)$ at the interface $x=0$ is applied although a barrier in an actual junction has finite thickness. 
Moreover, the spin-flip effect at the interface and the nonequilibrium properties of YBCO are neglected.
\par
Figure \ref{fig2} shows the temperature dependence of
conductance spectra of the junction.
As the temperature is lowered from room temperature,
no noticeable changes are detected above 70 K.
At temperatures below 40 K,
a gap-like structure (suppression of conductance) appears
at an energy level between $\pm$15 mV($\equiv \Delta$).
The presence of
the ZBCP becomes clear as the temperature is further decreased.
At the lowest temperature (4.2 K),
a large peak appears at zero-bias level.
The presence of the gap-like structure and ZBCP indicates that a barrier exists between the LSMO and YBCO layers.
In this study, however, we did not deposit any material as a barrier between the LSMO and YBCO layers.
It has been reported that normal metal/YBCO junctions in which no
insulating material had been deposited between the normal metal and YBCO layers
exhibited similar differential conductance spectra
\cite{lesueur_1,sanders_1}.
Thus, the barrier layer is expected to form naturally at the interface
between YBCO and other materials.
However, the composition and structure of the barrier in our junctions
are not clarified at present, because the barrier is too thin for these characteristics to be investigated.
Moreover, as mentioned in the introduction,
the existence of the ZBCPs is well explained in the tunneling
theory for anisotropic superconductors
by assuming that the tunneling current is governed by
in-plane ($ab$-plane) components \cite{tanaka_1,kashiwaya_2}.
In the cross-strip geometry, the in-plane contact between YBCO and LSMO existed at the side of YBCO film, 
and it also existed on the $c$-axis oriented surface 
because atomic force microscope measurement showed that it contains a large amount of $ab$-edges due to the island growth.
Therefore, the above assumption is reasonable
because the conductivity in the $ab$-plane is far larger than that for
the $c$-axis direction in the cases of high-$T_c$ superconductors.
The observed ZBCP is qualitatively
consistent with those found in other reports
on normal metal/YBCO tunneling junctions
\cite{kashiwaya_2,lesueur_1,covington_1,wei,sanders_1}.

On the other hand, the most significant difference between the present case
and those in the other reports
is that the counter electrode is not a normal metal but a ferromagnet,
and as a result the tunneling electrons are expected to be spin-polarized.
It has been theoretically shown that
the polarization can be estimated from the height of the ZBCP,
since the ZBCP is largely suppressed by the spin-polarization
\cite{kashiwaya_1,yoshida}.
This effect corresponds to the fact that the Andreev reflected
quasiparticle exists not as a propagating wave
but as an evanescent wave,
referred to as the virtual Andreev reflection process,
when the injection angle $\theta$ of quasiparticles
to the interface satisfies
$\sin^{-1}(k_{S}/k_{N,\uparrow})<\theta<\sin^{-1}(k_{N,\downarrow}/k_{N,\uparrow})$,
where $k_{S}$ is the Fermi-wave vector in superconductors.
Hence, the current through surface bound states
is prohibited when the energy of the quasiparticle
is less than the gap amplitude.
Figure~\ref{theory1} shows calculated normalized conductance spectra $\sigma(eV)$ for various polarizations.
($\sigma(eV)=\bar{\sigma}_S(eV)/\bar{\sigma}_N(eV)$,
where $\bar{\sigma}_S(eV)$ and $\bar{\sigma}_N(eV)$ are the tunneling conductance in superconducting and normal states, respectively.)
It is clear that the height of the ZBCP
is largely suppressed when the polarization becomes larger.
From this relationship,
it is estimated that the polarization
of the present experiments is much less than 90\%.
\par
Next, we discuss the effect of the applied magnetic
field.
Figure \ref{fig3} shows the
conductance spectra measured in various applied fields at 4.2 K.
As the magnetic field becomes larger,
an enhancement of the background conductance is always observed.
A similar feature has been reported by Vas'ko {\it et al.}
in a DyBa$_2$Cu$_3$O$_7$/La$_{2/3}$Ba$_{1/3}$MnO$_3$ junction \cite{vasko_2}, but has not been observed in normal metal/YBCO junctions \cite{kashiwaya_2,lesueur_1,covington_1,sanders_1}.
The origin of the change of background conductance will be discussed later.
All conductance spectra almost collapse onto a single curve except for a small change inside the gap
($|eV|<\mid \Delta\mid$) when they are
normalized by the conductance at $V=20$mV, as shown in
the inset of Fig.~\ref{fig3}.
On the other hand, a small field response can be seen
around zero-bias level.
Figure~\ref{fig4} shows the variation of
normalized conductance $\sigma^{\dagger}(eV,H)$ near zero-bias level
due to the applied field.
To clearly show the features,
the normalized conductance spectra are plotted with the zero field conductance
subtracted [$\sigma^{\dagger}(eV, H)-\sigma^{\dagger}(eV, 0)$] for several applied fields.
Two notable features have been observed.
One is the development of a dip at zero bias, indicating that the ZBCP splits into two peaks.
The ZBCP splitting has been observed
in normal metal/high-$T_c$ superconductor
junctions \cite{kashiwaya_2,lesueur_1,covington_1,sanders_1}.
The other is that the asymmetric heights of the
two shoulders are seen beside the dip.
This asymmetry may not be induced by the asymmetric background conductance,
because symmetric peak splitting has been observed in
normal metal/high-$T_c$ superconductor
junctions which exhibit asymmetric backgrounds
\cite{lesueur_1,sanders_1}.
Several possible origins of the ZBCP splitting have been proposed for $d$-wave
superconductors;
i) the Zeeman splitting of ${\pm g\mu_BH/2}$ in the energy levels
between up and down spins, where $g$ is the $g$-factor and $\mu_B$ is the Bohr
magneton,
ii) the inducement of the broken time-reversal symmetry (BTRS) states such
as ${d_{x^2-y^2}+is}$ wave \cite{covington_1,matsumoto_1,fogelstrom_1},
iii) the spin-filtering effect due to the ferromagnetic tunneling barrier
\cite{kashiwaya_1}.
In the following, we will show that the spin-filtering effect is the  most plausible
origin of the observed magnetic field response.
\par
The inset of Fig.~\ref{fig4} shows the amplitude of the peak splitting ${\delta_p}$
estimated from the peak-to-peak of the two
shoulders
as a function of applied magnetic field.
Completely different from usual Zeeman splitting
of which the response is linear to the applied field (${g\mu_BH/2}$),
${\delta_p}$ shows nonlinear behavior with respect to the applied field:
a rapid rise near the zero field ($H<0.5T$),
and almost linear behavior in the high field ($H>5T$).
Moreover, the observed ${\delta_p}$ is much larger than ${g\mu_BH}$
independent of $H$.
It is important to note that this behavior is  similar to
$M$--$H$ curves of a conventional paramagnet and is also consistent with
peak splitting due to the spin-filtering effect
which has been observed in Al/EuO/ and Al/EuS/ junctions \cite{hao_1,tedrow_1}.
Based on this fact, we can reject the possibility
of simple Zeeman splitting.
Moreover, although the inducement of BTRS can also explain the
nonlinear splitting,
the asymmetry in the peak splitting
observed in our result cannot be explained by this theory
as described in Ref.~\cite{kashiwaya_1}.
This is based on the fact that
the splitting due to the BTRS is not a spin-dependent effect.
\par
If we assume that the barrier naturally forms between YBCO and LSMO and that it attains ferromagnetic insulator nature similar to that of EuO and EuS barriers \cite{hao_1,tedrow_1}, 
the observed field response can be consistently explained in terms of the spin-filtering effect.
The strength of the exchange interaction
in the barrier (represented by ${\hat U}_{B}$ in Ref.~\cite{kashiwaya_1}) responds to the applied field in a similar way to paramagnetic materials \cite{note_2}.
In the applied magnetic field, the effective barrier height changes
between up and down spin components due to the finite ${\hat U}_{B}$,
then the tunneling electron begins to exhibit spin-dependent
energy splitting \cite{kashiwaya_1}.
Figure~\ref{theory2} simulates a theoretical calculation of magnetic field response of $\sigma(eV, H)-\sigma(eV,0)$ with $P$=30\%.
As $H$ becomes larger, the shoulders develop.
Moreover, the spin-polarization induces the
different peak splitting heights.
These features coincide with experimental data.
\par
The idea that the spin-filtering effect is the origin of ZBCP splitting is also supported by
the magnetic field response of the background conductance.
As described above, the background conductance increases
as the magnetic field is increased.
Although LSMO exhibits CMR at around the ferromagnetic transition temperature ($\sim$350 K),
the magnetoresistance of the LSMO film at 4.2 K was less than 1\% in the
measured magnetic field range.
Therefore, the increase of conductance with magnetic field originates in the transport property of the junction
rather than the change in the resistivity of LSMO.
In the framework of the spin-filtering effect,
the magnetic field response of background conductance is understood as the field dependence of $V_{\uparrow[\downarrow]}$:
i) in the absence of the field,
$V_{\uparrow}=V_{\downarrow}$ applies,
ii) as the magnetic field is increased,
$V_{\uparrow}$ decreases and $V_{\downarrow}$ increases,
iii) since the tunneling barrier height
for majority carrier (up-spin) decreases,
the total conductance of the junction is rapidly enhanced.
Figure~\ref{theory3} shows the simulated results for background conductance
as a function of $(V_{\downarrow}-V_{\uparrow})/(V_{\uparrow}+V_{\downarrow})$
for $P$=0\%, 30\%, and 60\% cases.
It is clear that as the difference in $V_{\downarrow}$ and $V_{\uparrow}$
increases, the background conductance also increases.
The influence of the polarization is  clear especially near the zero field.
Without the polarization, $\partial \sigma(eV)/\partial H$ is zero
near $H=0$.
This is because the effect of imbalance
in tunneling probabilities in up- and down-spins is
cancelled out without the polarization,
while the cancellation becomes smaller as the polarization
increases.
\par
We have shown that the observed conductance spectra and their
magnetic field response in LSMO/YBCO junctions
can be consistently understood in terms of the spin-filtering effect
in $d$-wave superconductors.
Two possibilities are deduced from the present results.
One is that a degraded layer existing between LSMO and YBCO
functions as an intrinsic barrier
and behaves as a ferromagnetic insulator.
The other is that a new type of magnetic boundary (surface) effect,
such as Shottokey barrier of spins,
exists at the interface or on the surface
due to the termination of the CuO or the MnO planes.
However, several questions still remain:
i) what kind of material exhibits the ferromagnetic insulator behavior at the YBCO/LSMO interface,
and ii) can the peak splitting observed in normal metal/high-$T_c$ superconductor junctions \cite{kashiwaya_2,lesueur_1,covington_1,sanders_1}
be attributed to the spin-filtering effect.
To clarify these problems, a more detailed characterization of the interface layer will be accomplished in the near future.
In addition, the origin of linear background conductance
has not been discussed here.
As is well known, this feature has been widely observed in
high-$T_c$ superconductor junctions
\cite{lesueur_1,alff_1,covington_1}.
Based on the above mentioned feature that the normalized conductance curves collapse
onto a single curve,
we assume that the origin of the linear background conductance is
an effect independent of the boundary properties discussed above.
Kirtley and Scalapino attributed
the origin of the energy dependent conductance
to an increment of tunneling probability
due to an inelastic tunneling process
via spin fluctuation \cite{kirtley_1}.
Although we believe that the present results do not contradict
this theory, further study is required to clarify this point.
\par

\section{Summary}
\label{conc}

We observed the magnetic field responses of the conductance spectra peculiar to LSMO/YBCO junctions, such as an increase of background conductance and asymmetric ZBCP splitting, which have not been observed in normal metal/YBCO junctions.
Moreover, the nonlinear response of $\delta_p$ to an applied field is different from the simple Zeeman splitting.
Although the inducement of BTRS can explain the nonlinear ZBCP splitting to applied field,
this explanation is not suitable because the asymmetry of the splitting cannot be explained.
On the other hand, it is shown that the observed features in the present study agree with the theory of tunneling spectroscopy for F/FI/d junctions
which assumes a spin dependent transmission (tunneling) probability
between a ferromagnet and a superconductor.
This suggests that the ferromagnetic barrier naturally forms between LSMO and YBCO,
and that the field response of the conductance spectra is due to the spin polarization of tunneling carriers and the spin-filtering effect.
From the present results, we deduce two possibilities of a ferromagnetic barrier.
One is that a degraded layer existing between LSMO and YBCO
functions as an intrinsic barrier
and behaves as a ferromagnetic insulator.
The other is that a new type of magnetic boundary (surface) effect,
such as Shottokey barrier of spins,
exists at the interface or on the surface
due to the termination of the CuO or the MnO planes.
\par

\acknowledgments
This work has been partially supported by the Core Research for Evolutional Science and Technology (CREST) of the Japan Science and Technology Corporation (JST) of Japan.



\begin{figure}
\caption{Schematic of LSMO/YBCO cross-strip type junction used for measurements.
\label{fig1}}
\end{figure}

\begin{figure}
\caption{Temperature dependence of  conductance spectra of LSMO/YBCO
junction.
ZBCP is clearly observed below 20 K.
\label{fig2}}
\end{figure}

\begin{figure}
\caption{Theoretical calculation of conductance spectra
$\sigma$
for F/FI/d junctions in the absence of an applied
magnetic field
as a function of polarization $P$.
In order to compare theory with experiments,
the smearing effect by energy of 0.3$\Delta$
is introduced in the calculation
and $V_{\downarrow}=V_{\uparrow}=3.5$ is assumed.
As $P$ becomes larger, ZBCP is suppressed.
\label{theory1}}
\end{figure}

\begin{figure}
\caption{Conductance spectra at 4.2 K as a function of applied magnetic field from 0 to 12 T.
Background conductance increases as magnetic field increases.
Inset shows normalized conductance spectra $\sigma^{\dagger}$. 
Each conductance spectrum is normalized by its conductance at $V=20$mV.
All the conductance spectra outside the gap almost collapse onto a universal curve and a small magnetic field response is observed around ZBCP.
\label{fig3}}
\end{figure}

\begin{figure}
\caption{Magnetic field dependence of the normalized conductance spectra around ZBCP. The zero-field conductance has been subtracted. 
Inset shows the peak splitting amplitude $\delta_p$ estimated from the peak-to-peak of two conductance shoulders.
Solid line represents the Zeeman splitting energy of ${g\mu_BH}$, where $g$=2.
\label{fig4}}
\end{figure}

\begin{figure}
\caption{Simulation of magnetic field dependence of $\sigma(eV, H)-\sigma(
eV,0)$ for
$P$=30\%.
The $H$ dependence of ${\hat U}_{B}$ is selected to reproduce the
experimental results.
Inset shows the case of $P$=0\%.
Asymmetric peak splitting is induced by finite polarization
in the ferromagnets.
\label{theory2}}
\end{figure}

\begin{figure}
\caption{Simulation of background conductance as a function of
$(V_{\downarrow}-V_{\uparrow})/(V_{\uparrow}+V_{\downarrow})$
for various $P$ (0, 30, 60 \%) values.
The increment of $(V_{\downarrow}-V_{\uparrow})/(V_{\uparrow}+V_{\downarrow})$
corresponds to enhancement of the exchange interaction
in the barrier (represented by ${\hat U}_{B}/{\hat V}_{0}$ in Ref. 24).
It is clear that the exchange interaction monotonically enhances the
background conductance.
The response to the magnetic field near the origin
becomes larger as the polarization become larger.
\label{theory3}}
\end{figure}


\begin{thebibliography}{99}
\bibitem{note_2} It is difficult to determine whether the observed response corresponds
to paramagnetic or ferromagnetic behavior because of the lack of
sensitivity near zero-field.
If the barrier is a ferromagnetic insulator,
hysteresis would be observed in the magnetic field response.
We did not observe such a trend.
\end{thebibliography}
\end{document}